\documentclass[]{iopart}
\usepackage{iopams}
\usepackage{setstack} 
\usepackage[]{graphicx}

\usepackage{color}
\usepackage[latin1]{inputenc}
\usepackage{amsfonts} 

\newcommand{\cU}{{\cal{E}}}
\newcommand{\bj}{{\bi{j}}}

\newcommand{\hu}{{\hat{\bi{u}}}}
\newcommand{\hE}{{\hat{\bi{E}}}}
\newcommand{\hj}{{\hat{\bi{j}}}}
\newcommand{\hr}{{\hat{\bi{r}}}}
\newcommand{\hz}{{\hat{\bi{z}}}}
\newcommand{\hh}{{\hat{\btheta}}}
\newcommand{\bl}{\bi{l}}

\begin{document}

\title[Continuous model of the coil-field interaction]{The watt balance operation: a continuous model of the coil interaction with the magnetic field}
\author{C P Sasso, E Massa and G Mana}
\address{INRIM -- Istituto Nazionale di Ricerca Metrologica, str.\ delle Cacce 91, 10135 Torino, Italy}

\begin{abstract}
In the watt balance experiments, separate measurements of the magnetic and electromotive forces in a coil in a magnetic field enable a virtual comparison between mechanical and electric powers to be carried out, which lead to an accurate measurement of the Planck constant. This paper investigates the three-dimensional nature of the coil-field interaction and describes the balance operation by a continuous three-dimensional model.
\end{abstract}





\section{Introduction}
Watt balances compare virtually the mechanical and the electric powers produced by the motion of a mass in the Earth gravitational field and by the motion of the supporting coil in a magnetic field \cite{Kibble:1976,Kibble:1987,Eichenberger:2003,Eichenberger:2009,Robinson:2009}. The electromotive force along a one-dimensional coil linking the magnetic flux $\Phi$ and moving at the velocity $u$ is $\cU=-(\partial_\hu \Phi)u$, where $\partial_\hu$ is the derivative along the motion direction. The component of the magnetic force acting on the same coil along the direction of motion, now carrying the electric current $I$, is $F_u=(\partial_\hu \Phi)I$. If this force counterbalances the weight $-mg$ of a mass $m$ in the gravitational field $g$, by combining these equations and eliminating the common factor $\partial_\hu \Phi$, we obtain $mgu=\cU I$. This equation relates virtually mechanical and electric powers and allows either $m$ to be determined in terms of electric quantities or the Planck constant to be determined in terms of mechanical quantities \cite{Mana:2012,Steiner:2013,Stock:2013,Mana:2013}.

A basic assumption in the previous analysis is that the coil wire is one-dimensional; hence, the three dimensional nature of a real wire deserves consideration. In the moving mode, the measured quantity is the difference between the voltage between the coil ends. In a one-dimensional coil, we identify this voltage with the electromotive force induced by the motion. In a three-dimensional coil, this is only possible if the Lorentz field $\bi{u}\times\bi{B}$, where $\bi{B}$ is the magnetic flux density, is conservative. In fact, only in this case there exists a scalar potential whose gradient is $\bi{u}\times\bi{B}$ and, therefore, there exists a charge distribution originating an electric field nullifying the Lorentz field. In general, the Lorentz field is not conservative and there is no electric field preventing eddy currents from flowing in the coil. This paper shows that, if $\bi{B}$ is radial and $\bi{u}$ is parallel to the field axis, $\bi{u}\times\bi{B}$ is conservative and the voltage between the coil ends is identical to the electromotive force.

In weighing mode, the magnetic force acting on a three-dimensional coil is given by the integral of the force density $\bi{j}\times\bi{B}$, where $\bi{j}$ is the density of the electrical current, over the coil volume. The current interaction with the magnetic field was investigated in \cite{Sasso:2013} and found to be insignificant. This paper investigates the inhomogeneity of the electric current due to the higher or lower resistance of the different current paths. It will be shown that, after modelling the coil turns as toroids, the current follows a $1/r$ law and the total force is calculable by a one-dimensional coil model.

This study was motivated by an unexplained spread -- the order of magnitude of which is 100 nW/W, to be compared with a targeted uncertainty of 10 nW/W -- of the Planck constant values reported by the International Avogadro Coordination (IAC) \cite{Andreas:2011a,Andreas:2011b}, the National Institute of Standards and Technology (NIST-USA) \cite{Steiner:2005,Steiner:2007}, the Swiss Federal Office of Metrology (METAS-Switzerland) \cite{Eichenberger:2011}, the National Physical Laboratory (NPL-UK) \cite{Robinson:2012a}, and the National Research Council (NRC-Canada) \cite{Steele:2012}.

\begin{figure}
\centering
\includegraphics[width=50mm]{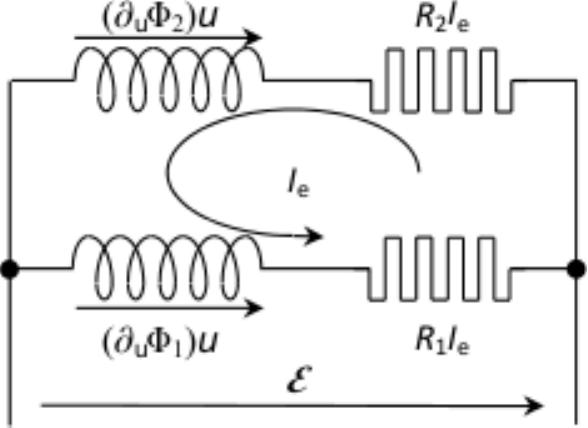}
 \hspace{10mm}
\includegraphics[width=51mm]{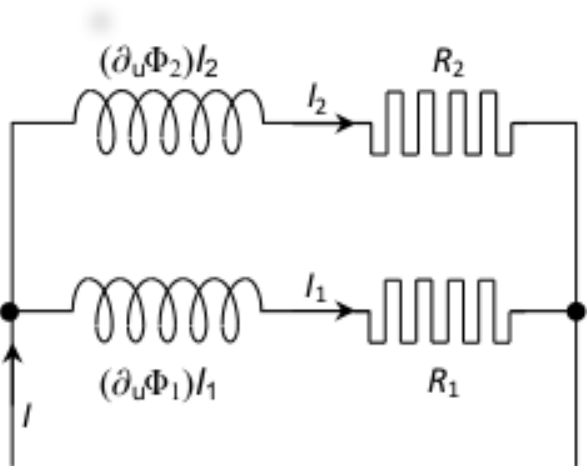}
\centering
\caption{Lumped parameter model of a three-dimensional coil. Left: moving mode. $I_e$ is the eddy current flowing in the mesh owing to the different electromotive forces, $(\partial_\hu \Phi_1)u$ and $(\partial_\hu \Phi_2)u$, induced by the coil motion. $R_1 I_e$ and $R_2 I_e$ are the voltage drops across the coil resistances. Right: weighing mode. $(\partial_\hu \Phi_1)I_1$ and $(\partial_\hu \Phi_2)I_2$ are the magnetic forces acting on the two coils.}\label{lumped}
\end{figure}

\section{Watt balance operation}
\subsection{One-dimensional model}
Let us start by considering a one dimensional loop $\gamma$ in a magnetic field having the magnetic flux density $\bi{B}$. Firstly, we consider the electromotive force induced -- in the absence of rotations -- by the coil motion with velocity $\bi{u}$,
\begin{equation}\label{moving}
 \cU = \oint_\gamma (\bi{u} \times \bi{B}) \cdot \rmd\bl = -(\partial_\hu \Phi)u ,
\end{equation}
where $u=|\bi{u}|$ and $\rmd\bl$ is the vector line element. A rotation-free motion implies that $\bi{u}$ is independent of the coordinates; hence, by the Stokes theorem and since $\bnabla\cdot\bi{B}=0$,
\begin{equation}
 \oint_\gamma (\hu \times \bi{B}) \cdot \rmd\bl = -\partial_\hu \Phi ,
\end{equation}
where $\Phi$ is the magnetic flux through the loop and $\partial_\hu=\hu\cdot\bnabla$ is the derivative along the velocity direction. Next, we assume that the electrical current $I$ flows along the same loop constrained to be at rest and calculate the component parallel to $\bi{u}$ of the magnetic force $\bi{F}$. Hence,
\begin{equation}\label{weighing}
 F_u = I \oint_\gamma \hu \cdot (\rmd\bl \times \bi{B}) = - I \oint_\gamma (\hu \times \bi{B}) \cdot \rmd\bl = (\partial_\hu \Phi)I .
\end{equation}
Since, in a one-dimensional coil, the difference $V$ between the electric potentials of the coil ends is the same as $\cU$, by combining (\ref{moving}) and (\ref{weighing}), we conclude that
\begin{equation}\label{meq}
 \bi{F}\cdot\bi{u} + V I = 0 .
\end{equation}
This is the measurement equation of the watt balance experiment.

In the present paper, we do not examine the full equation $\bi{F}\cdot\bi{u} + \bi{K}\cdot\bomega + V I = 0$, where $\bi{K}$ is the torque acting on the coil in the weighing mode and $\bomega$ is the coil angular-velocity in the moving mode. Furthermore, we do not examine the alignments necessary to identify the mechanical power $\bi{F}\cdot\bi{u}$ drained from or released by the moving coil with the power $m\bi{g}\cdot\bi{u}$ drained from or released by the motion of a mass $m$ in the gravitation field.

\subsection{Three-dimensional model}
The measurement equation (\ref{meq}) and the identity of the geometric factors multiplying the velocity and current in (\ref{moving}) and (\ref{weighing}) are consequences of the one-dimensional model of the coil. The magnetic force acting on a three-dimensional coil must be calculated by integrating the force density $\bi{j}\times\bi{B}$, where $\bi{j}$ is the density of the electrical current, over the coil volume. Therefore, in general, (\ref{weighing}) is not valid. Furthermore, the integration of the Lorentz field $\bi{u} \times \bi{B}$ along different paths joining the same endpoints gives, in general, different electromotive forces. Therefore, the difference between the electric potentials of the coil ends cannot be identified with (\ref{moving}).

\subsubsection{Lumped parameter model.}
If the coil is modelled by a number of one-dimensional wires connected in parallel and each wire is described by lumped parameters, the equation (\ref{meq}) can still be proven. For the sake of simplicity, let us consider only two coils connected in parallel as shown in Fig.\ \ref{lumped}. In the moving mode, the potential difference between the coil ends is
\numparts\begin{equation}\label{E-lumped}\fl
 V = (\partial_\hu \Phi_1) u - R_1 I_e = (\partial_\hu \Phi_2) u + R_2 I_e
 = \frac{(\partial_\hu \Phi_1)R_2 + (\partial_\hu \Phi_2)R_1}{R_1+R_2}\, u ,
\end{equation}
where the $I_e=(\partial_\hu \Phi_1-\partial_\hu \Phi_2)u/(R_1+R_2)$ is the eddy current flowing in the mesh and $R_1$ and $R_2$ are the coil resistances. In the weighing mode, the force acting on the two-coil system is
\begin{equation}\label{B-lumped}
 F_u = (\partial_\hu \Phi_1) I_1 +  (\partial_\hu \Phi_2) I_2 = \frac{(\partial_\hu \Phi_1)R_2 + (\partial_\hu \Phi_2)R_1}{R_1+R_2}\, I ,
\end{equation}\endnumparts
where $I=I_1+I_2$ is the measured current. Since the geometric factors in (\ref{E-lumped}) and (\ref{B-lumped}) are the same, the equation (\ref{meq}) is valid. In the following, we investigate if this identity is also valid when the continuous distributions of the resistance, current, and magnetic and electromotive forces are taken into account.

\begin{figure}
\centering
\includegraphics[width=70mm]{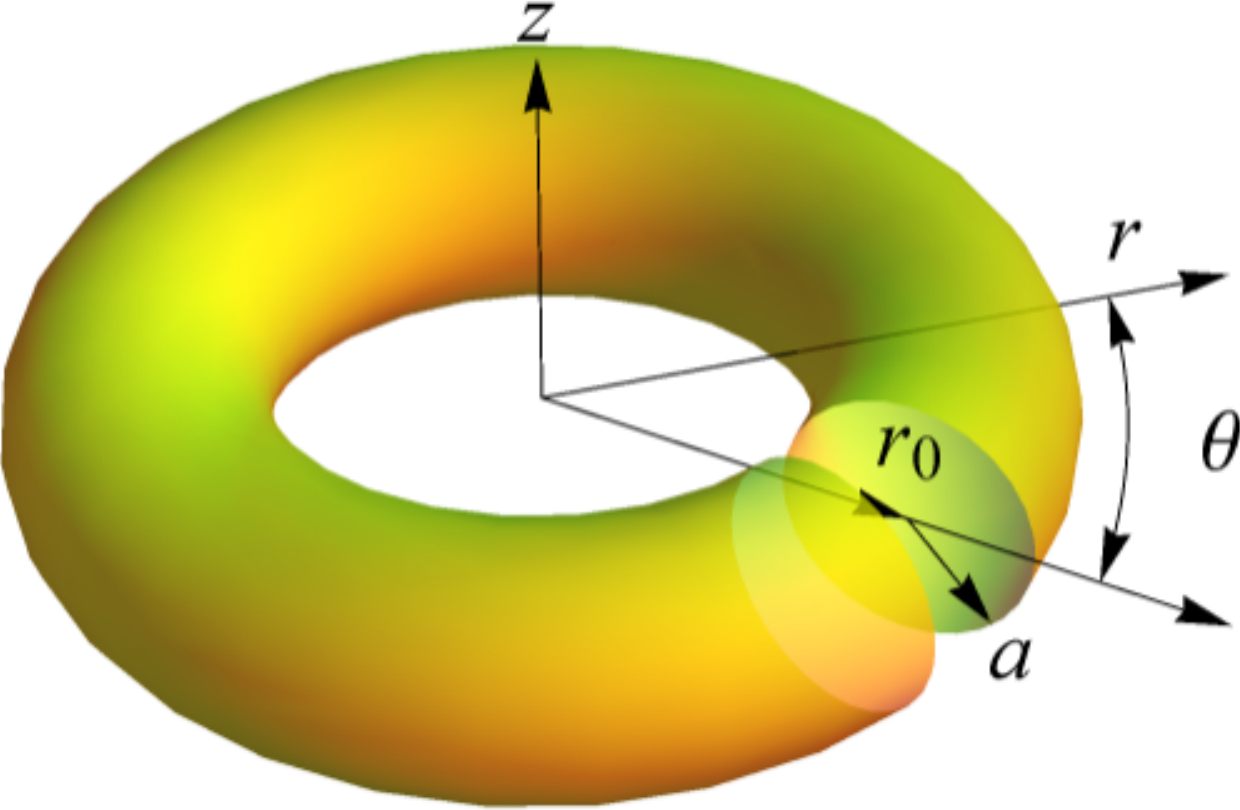}
\centering
\caption{Toroidal model of a watt-balance coil.}\label{ring}
\end{figure}

\section{Coil-field interaction}
\subsection{Current distribution}
As shown in Fig.\ \ref{ring}, we consider a model where each coil turn is a split toroid of radius $r_0$ and circular cross section of radius $a$ -- but the results are independent of the actual cross section. The toroid end is figured joined to the next toroid start, so that a full revolution carries us in the next turn of the coil. To find the current distribution, we neglect the magnetic field, whose effect was investigated in \cite{Sasso:2013}, and use the cylindrical polar coordinates $r$, $\theta$, and $z$, with the origin fixed at the toroid centre and the $z$ axis tied to the toroid axis.

\begin{figure}
\centering
\includegraphics[width=60mm]{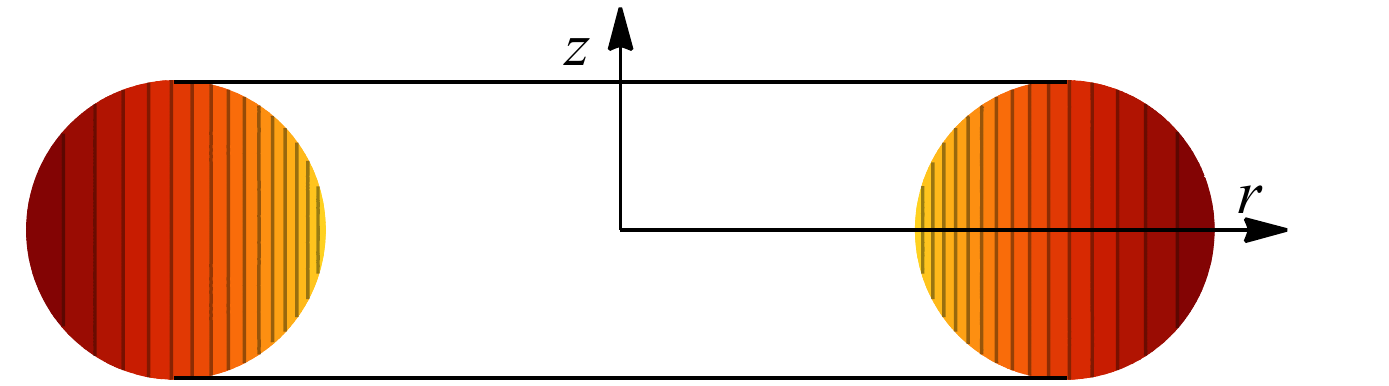}
\centering
\caption{Distribution of the electric current in the toroid model of a watt-balance coil, shown in cross section. The brighter colour indicates the higher current density on the inner path.}\label{section}
\end{figure}

Inside the coil, the electric potential $\phi$ satisfies the Laplace equation
\begin{equation}\label{Laplace}
 \Delta\phi=0 ,
\end{equation}
with the Neumann boundary conditions $(\bnabla\phi)\cdot\bi{n}=0$, where $\bi{n}$ is the unit normal to the toroid surface. These boundary conditions ensure that no electric current flows through the coil surface and that $\phi=-r_0 E_0 \theta$, where the $r_0E_0$ factor is chosen for later convenience, is the unique solution of (\ref{Laplace}), up to a meaningless additive constant. Hence, the electric field is given by
\begin{equation}\label{field}
 \bi{E} = -\bnabla\phi = \frac{r_0 E_0}{r} \,\hat{\btheta} ,
\end{equation}
where $\hat{\btheta}$ is the polar unit vector. According the Ohm law, the electric-current density is
\begin{equation}\label{current}
 \bi{j} = \sigma \bi{E} = \frac{r_0 j_0}{r} \,\hat{\btheta} ,
\end{equation}
where $\sigma$ is the coil conductivity and $j_0=j(r_0)=\sigma E_0$ is the current density evaluated at $r=r_0$. A contour plot of the electric current density is shown in Fig.\ \ref{section}; according to (\ref{current}), on the inner path the current density is the highest.

\subsection{Magnetic field}
Let the coil be in a magnetic field whose associated flux density is
\begin{equation}\label{B:field}
 \bi{B} = \frac{r_0B_0}{r} \,\hat\bi{r} ,
\end{equation}
where $\hat\bi{r}$ is the radial unit vector and $\hz$ is the field axis. The $r_0B_0$ factor does not have any particular meaning, but $\bi{B}(r_0)=B_0\hat\bi{r}$. It must be noted that (\ref{B:field}) is singular at $r=0$. Therefore, a region about the field axis must be removed from the $\bi{B}$ domain and, consequently, we cannot evaluate the magnetic and electromotive forces by means of the derivative of linked flux.

\subsection{Measurement equation}
We consider a coaxial system, where the coil and field axes coincide and prove that, provided the coil moves along the field axis, without parasitic rotations and with velocity $\bi{u}=u\hat\bi{z}$, the equation (\ref{meq}) is valid.

\subsubsection{Moving mode.}
The Lorentz field originated by the coil motion,
\numparts\begin{equation}
 \bi{u} \times \bi{B} = \frac{u r_0 B_0}{r}\, \hat{\btheta} ,
\end{equation}
where $\bi{u}=u\hat\bi{z}$, is counterbalanced by the electric field
\begin{equation}
 \bi{E} = -\frac{u r_0 B_0}{r}\, \hat{\btheta} ,
\end{equation}\endnumparts
originated by opposite charges at the coil ends \cite{Sasso:2013}; a detailed proof of this assertion will be given in section \ref{subsub-moving-mode}. Therefore, the difference between the electrical potentials of the coil ends,
\begin{equation}\label{V}
 V = -\int_0^{2\pi} \hspace{-2mm} (\bi{E}\cdot\hat{\btheta})\, r\,\rmd\theta = BLu ,
\end{equation}
where $L=2\pi r$ is the circle length, $B=r_0 B_0/r$, and $BL=B_0L_0$ is a constant, is a well defined quantity.

\subsubsection{Weighing mode.}
To prove that, if the coil moves along the field axis, the equation (\ref{meq}) is valid, we must calculate the mechanical power drained from or released by a coil moving with velocity $\bi{u}=u\hat\bi{z}$. Hence,
\begin{equation}\label{static:1}
 \bi{F}\cdot\bi{u} = - \int_0^{2\pi} \int_{r_-}^{r_+} \int_{z_-}^{z_+} \hspace{-2mm}
 \bi{u} \cdot (\bi{B} \times \bi{j}) \, r\, \rmd z\, \rmd r\, \rmd \theta ,
\end{equation}
where $r_{\pm}=r_0 \pm a$ and $z_\pm=\pm\sqrt{a^2-(r-r_0)^2}$ (see Fig.\ \ref{ring}). By using (\ref{current}) and (\ref{B:field}) in (\ref{static:1}) and by observing that $\hat\bi{r} \times \hat{\btheta} = \hat\bi{z}$, we obtain,
\begin{equation}\label{static:2}
 \bi{F}\cdot\bi{u} = -2\pi r_0 j_0 \left( r_0-\sqrt{r_0^2-a^2} \right) BL u = -V I ,
\end{equation}
where we used (\ref{V}) and
\begin{equation}\label{tot_current}
 I = \int_{r_-}^{r_+} \int_{z_-}^{z_+} \hspace{-2mm}
 (\bi{j} \cdot  \hat{\btheta})\, \rmd z\, \rmd r =
 2\pi r_0 j_0 \left( r_0-\sqrt{r_0^2-a^2} \right)
\end{equation}
is the current flowing in the coil. It must be noted that, in the limit when $r_0 \rightarrow \infty$, (\ref{tot_current}) reduces to $I=\pi a^2j_0$.

\section{Geometric aberrations}
This section investigates the effect of the coil imperfections, such as misplacements, tilts, and shape errors. Apart from these errors, the field is assumed radial and the motion is assumed parallel to the field axis and free from parasitic rotations.

\subsection{Moving mode.}\label{subsub-moving-mode}
In a radial field, provided that $\bi{u}$ is parallel to the field axis, $\bi{u} \times \bi{B}$ is a conservative field and the electromotive force is the same along any loop circling the same number of times the axis and it is null if the loop does not enclose the axis, no matter what the loop shape may be.

To prove this assertion, let us write both the field and coil velocity in the orthonormal basis $(\hat{\bi{r}}, \hat{\btheta}, \hz)$ of the cylindrical polar chart and let us set $\hz$ parallel to the field axis. In the general case, if its Cartesian components are $u_x$, $u_y$, and $u_z$, the cylindrical components of the velocity are
\begin{equation}\label{velocity}
 \bi{u} = u_r \hat{\bi{r}} + u_\theta \hat{\btheta} + u_z \hz ,
\end{equation}
where $u_r=u_x \cos(\theta)+u_y \sin(\theta)$ and $u_\theta=u_y \cos(\theta)-u_x \sin(\theta)$. Hence, the Lorentz field is
\begin{equation}\label{L-unit}
 \bi{u} \times \bi{B} = (u_z \hat{\btheta} - u_\theta \hz)B ,
\end{equation}
where it has been used (\ref{B:field}) and $B=r_0 B_0/r$. Since, the vector line element of an arbitrary curve $\gamma$ -- represented by the parametric equations $r=r(\tau)$, $\theta=\theta(\tau)$, and $z=z(\tau)$ and parameterised by $\tau$ -- is
\begin{equation}
 \rmd \bl = \left[ (\partial_\tau r) \hat{\bi{r}} + r (\partial_\tau \theta) \hat{\btheta} + (\partial_\tau z) \hz \right] \rmd\tau ,
\end{equation}
the infinitesimal electromotive force is
\begin{equation}\label{dE}
 \rmd \cU = (\bi{u} \times \bi{B})\cdot \rmd\bl = \left[ r (\partial_\tau \theta) u_z - (\partial_\tau z) u_\theta \right] B\, \rmd\tau .
\end{equation}

When $u_\theta=0$, that is, when the velocity is parallel to the field axis, (\ref{dE}) reduces to
\begin{equation}
 \rmd \cU = u r B \,\rmd\theta ,
\end{equation}
where $u=u_z$, $(\partial_\tau \theta)\rmd\tau=\rmd\theta$, and $rB=r_0B_0$ is a constant. Eventually, the electromotive force,
\begin{equation}\label{E-gamma}
 \cU = \int_\gamma (\bi{u} \times \bi{B} )\cdot \rmd\bl = u r B \int_0^{2\pi} \hspace{-2mm} \rmd\theta = B L u ,
\end{equation}
where $L=2\pi r$ and $BL=B_0 L_0$ is a constant, is the same along any loop $\gamma$ encircling the axis.

A complementary way to prove (\ref{E-gamma}) is to observe that, if $\bi{u}$ is parallel to the axis of a radial field, $\bi{u} \times \bi{B}=\bnabla\phi$, where the potential function is $\phi=r_0 B_0 \theta$. Hence, the electric charges can distribute inside the coil to originate the electric field $\bi{E}=-\bnabla\phi$ nullifying $\bi{u} \times \bi{B}$, preventing eddy currents from flowing in the coil, and giving raise to an unique voltage,
\begin{equation}\label{VeqE}
 V = -\int_\gamma \bi{E} \cdot \rmd\bl = \int_\gamma (\bi{u} \times \bi{B} )\cdot \rmd\bl = \cU ,
\end{equation}
between the coil ends. It must be noted that the coil ends are assumed to be equipotential surfaces, that is, $\theta=$ const.\ on the coil end-surfaces.

\subsection{Weighing mode.}
If $\bi{j}$ is constant along the flow lines -- that is, if $\partial_\hj j=0$, where $\partial_\hj$ is the derivative along the $\bi{j}$ direction -- a comprehensive study of the magnetic force acting on an arbitrarily shaped three-dimensional coil is possible. There is no general rule to establish if this is likely to be the case, but a couple of considerations matter. Owing to the Ohm's law, $\bi{j}=\sigma\bi{E}$, where we neglect the Lorentz force due to the current interaction with the magnetic field \cite{Sasso:2013}, the $\partial_\hj j=0$ assumption is equivalent to $\partial_\hE E=0$. In addition, the continuity equation, $\bnabla\cdot\bi{j}=0$, and $\partial_\hj j=0$ enforce a constant cross-section of the coil wire.

If $\partial_\hE j=0$, the magnitude of the current density, $|\bi{j}|=j(\xi,\eta)$, is only a function of the $\xi$ and $\eta$ coordinates in the equipotential surface $\Sigma$ orthogonal to $\bi{j}$. Therefore, the power drained from or released by a coil motion parallel to the field axis is
\begin{eqnarray}\label{F-gamma}
 \bi{F}\cdot\bi{u} &= &- \int \hspace{-2mm} \int_\Sigma j \left[ \oint_{\gamma(\xi,\eta)} \hspace{-5mm} \bi{u}\cdot(\bi{B} \times \rmd\bl) \right] \rmd\xi\, \rmd\eta
 \\ \nonumber
 &= &- \int \hspace{-2mm} \int_\Sigma j \left[ \oint_{\gamma(\xi,\eta)} \hspace{-5mm} (\bi{u} \times \bi{B} )\cdot \rmd\bl \right] \rmd\xi\, \rmd\eta
 = -VI ,
\end{eqnarray}
where $\bi{u}=u\hat\bi{z}$ is the coil velocity, $\rmd\xi\, \rmd\eta$ is the area element, $\rmd\bl$ is the vector line element along the $\gamma(\xi,\eta)$ current line, the outermost integration is carried out on all the current lines,
\begin{equation}\label{oint}
 \oint_{\gamma(\xi,\eta)} \hspace{-5mm} (\bi{u} \times \bi{B} )\cdot \rmd\bl = V
\end{equation}
is independent of the integration line (see section \ref{subsub-moving-mode}), and
\begin{equation}
 I = \int \hspace{-2mm} \int_\Sigma j(\xi,\eta)\, \rmd\xi\, \rmd\eta
\end{equation}
is the current flowing in the coil. Equations (\ref{VeqE}) and (\ref{F-gamma}) are central to our analysis. They show that, in a radial field, the equation $\bi{F}\cdot\bi{u} + VI = 0$ holds, no matter what the shape, placement, and tilt of the coil may be.

\section{Non axial motion}
This section investigates a coil motion that is not parallel to the field axis. The field and coil are still assumed coaxial and the motion free from parasitic rotations. In this case, the Lorentz field is
\begin{equation}\label{uB}
 \bi{u} \times \bi{B} = \left\{ u_z \hh + \left[ u_x \sin(\theta) - u_y \cos(\theta) \right] \hz \right\} \frac{r_0 B_0}{r} ,
\end{equation}
where we used cylindrical polar coordinates tied to the field and coil axis, the origin is in the coil centre, the coil velocity and the field are given by (\ref{B:field}) and (\ref{velocity}), respectively. With the assumption of a one-dimensional coil, the effect of an imperfect motion is discussed in \cite{Robinson:2012}.

\subsection{Moving mode}\label{m-tilt}
\begin{figure}
\centering
\includegraphics[width=70mm]{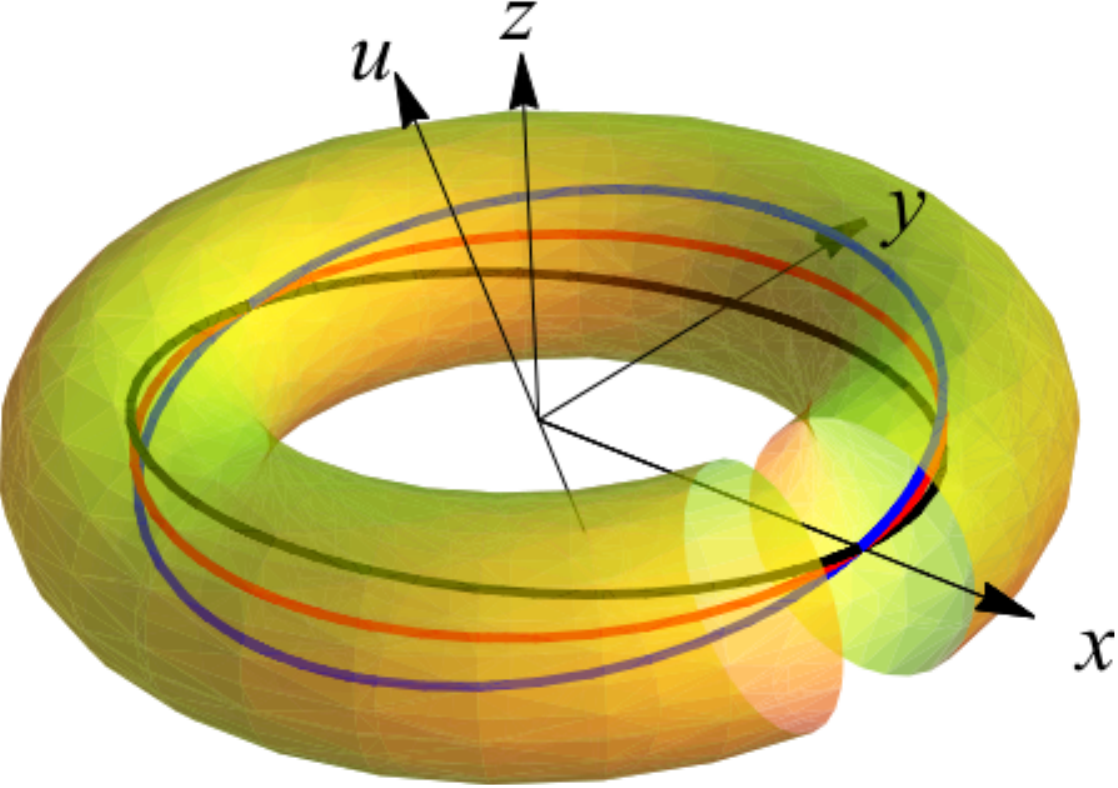}
\centering
\caption{Non-axial coil motion: the electromotive force along the sheaf's circles depends on the circle tilt. $z$ is the field and coil axis; $\bi{u}$ is the coil velocity.} \label{sheaf}
\end{figure}

Let us consider a sheaf of tilted circles -- centred in the origin and having radius $R$ (see Fig.\ \ref{sheaf}). The parametric equations of any sheaf's circle are
\begin{equation}\label{c:tilted}
 r \approx R, \; \theta \approx \tau, \; z \approx \alpha R \sin(\tau),
\end{equation}
where $0\le \tau < 2\pi$ is the circle's azimuth, $\alpha \ll 1$ is the tilt, the $x$ axis has been set parallel to the tilt axis, and only the leading -- first order -- terms of the series expansions (if not null) have been retained. The vector line element along (\ref{c:tilted}) is
\numparts\begin{equation}
 \rmd\bl \approx R \left[ \hh +\alpha\cos(\tau) \hz \right]\rmd\tau .
\end{equation}
The field,
\begin{equation}
 \bi{B} \approx \frac{r_0 B_0}{R} \hr ,
\end{equation}
and coil velocity,
\begin{equation}
 \bi{u} \approx \left[u_x \cos(\tau)+u_y \sin(\tau)\right] \hat{\bi{r}} + \left[u_y \cos(\tau)-u_x \sin(\tau)\right] \hat{\btheta} + u_z \hz ,
\end{equation}\endnumparts
values in the circle points are similarly approximated by neglecting the second and higher order terms.

The electromotive forces along the sheaf's circles,
\begin{eqnarray}\fl \label{E-tilt}
 \cU(\alpha) &= &\oint ( \bi{u} \times \bi{B} )\cdot \rmd\bl = r_0 B_0 \int_0^{2\pi} \left\{ u_z +
 \alpha\cos(\tau)\left[ u_x\sin(\tau) - u_y\cos(\tau) \right] \right\}\, \rmd\tau \\ \nonumber \fl
 &= &B_0 L_0 u_z\left( 1 - \frac{\alpha u_y}{2u_z} \right) ,
\end{eqnarray}
where $L_0=2\pi r_0$, depend on the circle tilt $\alpha$. Since, the sheaf's circles intersect in the same points -- $(r=R,\theta=0,z=0)$ and $(r=R,\theta=\pi,z=0)$ -- (\ref{E-tilt}) proves that the field (\ref{uB}) is not conservative. Therefore, there is not any electric field -- which is conservative -- nullifying $\bi{u} \times \bi{B}$ everywhere and eddy currents flow in the coil.

\subsection{Weighing mode}
In the weighing mode, since the current lines are horizontal circles centred on the field axis, the power drained from or released by the coil motion with velocity $\bi{u}$ is given by line integral
\begin{equation}\label{tilted-meq}
 \bi{F}\cdot\bi{u} = -I \oint (\bi{u}\times\bi{B}) \cdot \rmd\bl = -\cU_0 I ,
\end{equation}
where we used (\ref{F-gamma}) and (\ref{E-tilt}) and $\cU_0 =\cU(\alpha=0)$. However, $\cU_0$ has not been proven or disproved to be the voltage between the coil ends as measured in the moving mode and, therefore, (\ref{tilted-meq}) does not prove or disprove the measurement equation (\ref{meq}). To solve this issue requires that the eddy currents are reckoned in the calculation of the difference between the electric potentials of the coil ends. The section \ref{eddy-currents} will outline how these currents are included in the model of the balance operation.

\section{Non-radial field}
The previous section showed that the $1/r$ law ensures that the equation (\ref{meq}) holds, no matter what the coil shape and position may be.
Deviations from the $1/r$ law make the line integral (\ref{oint}) dependent on the path $\gamma$, which makes factorisation of (\ref{F-gamma}), and hence
identification with $\partial_\hu \Phi$ in (\ref{moving}) and (\ref{weighing}), impossible. Therefore, this section investigates the operation of a toroidal coil in a non-radial, but axially symmetric, field. The coil and field axes are still assumed coaxial and we limit the analysis to a motion parallel to the field axis.

\subsection{Magnetic field}
Let
\begin{equation}\label{aberrated0}
 \bi{B} = \left[ \frac{B_0 r_0}{r} + f(r,z) \right] \hat{\bi{r}} - h(z)\, \hat\bi{z} ,
\end{equation}
where $f(r_0,z)=0$, be the flux density. Since, $\bi{B}$ must satisfy $\bnabla\cdot\bi{B}=0$ and $\bnabla\times\bi{B}=0$, the functions $f(r,z)$ and $h(z)$ are linked by
\begin{equation}
 f = \frac{(r^2-r_0^2)\partial_z h}{2r} .
\end{equation}
For example, with the simplest choice $h(z)=-\lambda z$ describing a gradient of the vertical field-component, the flux density is
\begin{equation}\label{aberrated1}
 \bi{B} = \frac{2B_0 r_0+\lambda(r^2-r_0^2)}{2r}\, \hat{\bi{r}} - \lambda z\, \hat\bi{z} .
\end{equation}

\subsection{Weighing mode}
By using (\ref{current}) and (\ref{aberrated0}) in (\ref{static:1}), the component of the magnetic force along the field axis is
\begin{equation}\label{aberratedF0}
 F_z = -2\pi r_0 j_0 \int_{r_-}^{r_+} \int_{z_-}^{z_+} \hspace{-2mm} B_r(r)\, \rmd z\, \rmd r ,
\end{equation}
where $B_r(r)$ is the radial component of (\ref{aberrated0}). To make an example, by using (\ref{aberrated1}) in (\ref{aberratedF0}), we obtain
\begin{equation}\fl\label{aberratedF1}
 F_z = -BLI \left[ 1 + \frac{r_0 \big(\varrho^2/2 - 1 +\sqrt{1-\varrho^2}\big) \lambda}{2B_0 \big(1-\sqrt{1-\varrho^2}\big)} \right]
 \approx -BLI \left( 1 - \frac{\varrho^2 \lambda r_0}{8B_0} \right) ,
\end{equation}
where $B=B_0 r_0/r$, $L=2\pi r$, and $\varrho=a/r_0 \ll 1$.

\subsection{Moving mode}
By using (\ref{aberrated1}) in
\begin{equation}
 \cU = \int_0^{2\pi}( \bi{u} \times \bi{B} ) \cdot \hat{\btheta} \, r\, \rmd \theta ,
\end{equation}
it follows that the electromotive force induced by the coil motion along a horizontal circle centred on the coil axis,
\begin{equation}\label{Er}
 \cU(r) = 2\pi r B_r(r) u = BLu \left[ 1 + \frac{\lambda(r^2 - r_0^2)}{2r_0 B_0} \right] ,
\end{equation}
where the velocity $\bi{u}=u\hz$ is parallel to the field axis, depends on the circle radius $r$. A naive way to find a unique difference between the potentials of the coil ends is to average (\ref{Er}). The result is
\begin{equation}\label{Er-mean}
 V = \overline{\cU} = \frac{2u}{a^2} \int_{r_-}^{r_+} \int_{z_-}^{z_+} \hspace{-2mm} r B_r(r)\, \rmd z\, \rmd r
 \approx BLu \left( 1 + \frac{\varrho^2 \lambda r_0}{8B_0} \right) .
\end{equation}

\begin{figure}
\centering
\includegraphics[width=50mm]{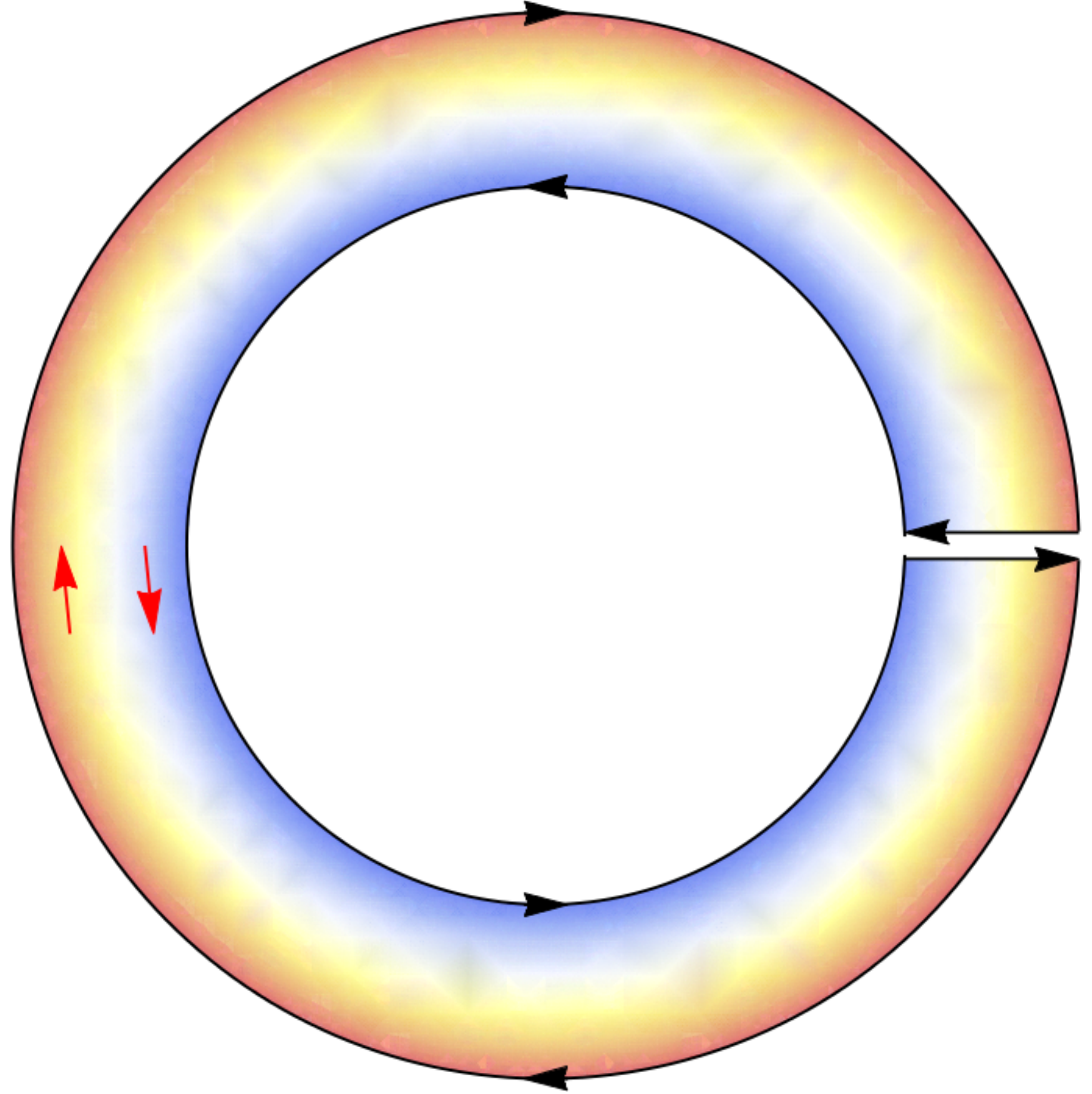}
\caption{Eddy current density in the $z=0$ section of the coil. The current flows counter clockwise (blue) in the inner paths, clockwise (red) in the outer paths. In the middle (white) the current density is zero. The eddy current comes from the previous turn of the coil and, after a full revolution, flows in the next turn. The black arrows indicate the circuit used to calculate the electromotive force (\ref{Er2}).}\label{loop}
\end{figure}

However, this averaging procedure is not really sound. Firstly, we averaged on the horizontal circles centred on the field axis, but there are infinite paths joining the coil ends that have not been included in the average. Secondly, the electromotive force along the circuit shown in Fig.\ \ref{loop},
\begin{equation}\label{Er2}
 \cU_\circ = \cU(r_+) - \cU(r_-) = 2a\lambda L_0 u ,
\end{equation}
is not zero and we did not take the effect of the eddy current induced by $\cU_\circ$ into account. The electromotive force $\cU_\circ$ is due to the gradient $-\partial_z \Phi=2a\lambda L_0$ of the flux $\Phi=2aLB_z=-2a\lambda L z$ linked to the circuit. As shown Fig.\ \ref{loop}, $\cU_\circ$ drives opposite eddy currents along the inner and outer parts of the coil.

The next section will remedy these weaknesses. For the moment, let us note that, by combining (\ref{aberratedF1}) and (\ref{Er-mean}), we obtain a corrected measurement equation
\begin{equation}\label{aberrated-meq}
 \bi{F}\cdot\bi{u} + V I \left( 1 - \frac{\varrho^2 \lambda r_0}{4B_0} \right) = 0 .
\end{equation}
To give an order-of-magnitude estimate of the correction, we use $a \approx 0.2$ mm, $r_0 \approx 100$ mm, $B_0=0.5$ T, and $\lambda \approx 10^{-6}$ mm$^{-1}$. Hence, the correction to be applied is about 0.8 nW/W. When compared to the typical 30 nW/W uncertainty associated to the watt balance measurements, this value is reassuringly small.

\subsubsection{Continuous model of the eddy current.}\label{eddy-currents}
Owing to the Lorentz field $\bi{u}\times\bi{B}$, the charges inside the coil redistribute and originate a counteracting electric field $\bi{E}$. Therefore, the eddy current density is given by
\begin{equation}\label{eddy}
 \bi{j} = \sigma ( \bi{E} + \bi{u}\times\bi{B} ) .
\end{equation}
This is the Ohm law for a moving coil. Since no current flows through the coil surface, $\bi{j}\cdot\bi{n}=0$, where $\bi{n}$ is the unit normal to the interface. Hence, in general, the boundary conditions of the electric potential are
\begin{equation}\label{BC}
 (\bnabla\phi)\cdot\bi{n} = (\bi{u}\times\bi{B})\cdot\bi{n} ,
\end{equation}
where $\bnabla\phi=-\bi{E}$. In the specific case we are considering -- by neglecting the end surfaces, which is equivalent to consider an infinite coil -- since $(\bi{u}\times\bi{B})\cdot\bi{n}=0$, (\ref{BC}) simplifies as $(\bnabla\phi)\cdot\bi{n} = 0$. Therefore, the sought field is the same as (\ref{field}), where $E_0$ is determined in such a way that
\begin{equation}\label{Tot-I}
 \int_{r_-}^{r_+} \int_{z_-}^{z_+} (\bj\cdot\hat{\btheta})\, \rmd z\, \rmd r = 0 ,
\end{equation}
because the total current is zero. Hence, by combining (\ref{eddy}) and (\ref{Tot-I}), we obtain the identity
\begin{equation}\label{ide}
 r_0 E_0 \int_{r_-}^{r_+} \int_{z_-}^{z_+} \rmd z\, \rmd r /r = -u \int_{r_-}^{r_+} \int_{z_-}^{z_+} B_r(r)\, \rmd z\, \rmd r .
\end{equation}
Eventually, the difference between the electric potentials of the coil ends is
\begin{equation}\label{Er-eddy}
 V = -\int_0^{2\pi} \hspace{-2mm} (\bi{E}\cdot\hat{\btheta})\, r\rmd\theta =
 -2\pi r_0 E_0 = \frac{2\pi u \displaystyle\int_{r_-}^{r_+} \hspace{-2mm} \int_{z_-}^{z_+} B_r(r)\, \rmd z\, \rmd r}
 {\displaystyle\int_{r_-}^{r_+}  \hspace{-2mm} \int_{z_-}^{z_+} \rmd z\, \rmd r /r} ,
\end{equation}
were we used (\ref{ide}). By observing that, from (\ref{aberratedF0}) and (\ref{current})
\numparts\begin{equation}
 \int_{r_-}^{r_+} \hspace{-2mm} \int_{z_-}^{z_+} \hspace{-1mm} B_r(r)\, \rmd z\, \rmd r = -\frac{F_z}{2\pi r_0 j_0}
\end{equation}
and
\begin{equation}
 r_0 j_0 \displaystyle\int_{r_-}^{r_+} \hspace{-2mm} \int_{z_-}^{z_+} \hspace{-1mm} \rmd z\, \rmd r /r = I ,
\end{equation}\endnumparts
we can rewrite (\ref{Er-eddy}) as $\bi{F}\cdot\bi{u} + V I = 0$. Therefore, provided field is axially symmetric, the coil is a toroid sitting in the same plane as the field, and the motion occurs along the field axis, the measurement equation (\ref{meq}) holds, no matter what the radial profile of the field may be.

\section{Conclusions}
The theory of the watt-balance operation rests on a lumped parameter model of the coil-field interaction, where the coil is considered one-dimensional. We investigated if the theory's results are still valid when the three-dimensional nature of the coil is taken into account and the balance operation is described by continuous parameters.

Since the magnetic force is given by an integral over the coil volume and, in general, the electromotive forces along different paths joining the same endpoints are different, we cannot give a general proof of the $\bi{F}\cdot\bi{u} + V I = 0$ relationship between the force and current measured in the weighing mode and the velocity and voltage measured in the moving mode. However, we did not find a counter example.

The $\bi{F}\cdot\bi{u} + V I = 0$ equation has been proven when a toroidal coil is coaxially placed in a radial magnetic field and moved along the field axis. Furthermore, it has been also proven valid when a number of aberrations of the coil-field interaction are considered one at a time, with the only case of a coil motion that is not parallel to the field axis remaining undecided. Therefore, small deviations from an experimental set-up where a toroidal coil is coaxially placed in a radial field and moved along the field axis cause second order errors, in the worst case.

\ack
This work was jointly funded by the European Metrology Research Programme (EMRP) participating countries within the European Association of National Metrology Institutes (EURAMET) and the European Union. We thank a referee of the initial version of this paper, whose criticisms prompted this detailed investigation, and Ian Robinson, whose hints about the lumped parameter model helped to clarify the role of the eddy currents.

\end{document}